\documentclass[a4paper,onecolumn]{article}

\usepackage{supertabular,booktabs}

\usepackage[english]{babel}
\usepackage[utf8]{inputenc}
\usepackage{graphicx}  
\usepackage{latexsym}  
\usepackage{amssymb,amsmath,amsthm,amsfonts} 
\usepackage{color} 
\usepackage{epstopdf}
\usepackage{enumerate}
\usepackage{widetext}
\usepackage{tensor}

\usepackage{tikz}
\usepackage{pgfplots}
\usepackage{xcolor}

\definecolor{findOptimalPartition}{HTML}{D7191C}
\definecolor{storeClusterComponent}{HTML}{FDAE61}
\definecolor{dbscan}{HTML}{ABDDA4}
\definecolor{constructCluster}{HTML}{2B83BA}

\usepackage[margin=0.5in]{geometry}
\usepackage{authblk}
\usepackage[pdfborder={0 0 0}]{hyperref}

\usepackage{float}
\usepackage[caption = false]{subfig}

\usepackage{lipsum}
\usepackage{multicol}

\usepackage[normalem]{ulem} 

\usepackage{color, colortbl}
\definecolor{Gray}{gray}{0.9}
\definecolor{Red1}{rgb}{0.9,0.57,0.1}
\definecolor{Red2}{rgb}{0.85,1,.9}
\definecolor{Cyan}{rgb}{0.88,1,1}

\usepackage{pgfplots,filecontents}
\usepackage{siunitx}

\begin{document}

\title{Cosmology from a Lagrangian formulation for Rastall's theory}


\renewcommand\Authands{ and }


\author[1]{Renato Vieira dos Santos}
\author[2]{Jos\'{e} A. C. Nogales}

\affil[1]{UFLA - Universidade Federal de Lavras\\
DFI - Departamento de Física, CEP: 37200-000, Lavras, Minas Gerais, Brazil.\\

\url{renato.santos@dfi.ufla.br} \& \url{econofisico@gmail.com}}

\affil[2]{UFLA - Universidade Federal de Lavras
DFI - Departamento de Física, CEP: 37200-000, Lavras, Minas Gerais, Brazil.\\

\url{jnogales@dfi.ufla.br}}

%

\date{\today}

\vspace{1cm}

    \maketitle
    \begin{abstract}
We give a Lagrangian formulation for the theory of Rastall of gravitation. After proposing a Lagrangian density that reproduces the equations of motion postulated by Rastall, we study the cosmological consequences and fit the parameters using recent data from Hubble function $H(z).$ According to two model selection criteria, one based on corrected Akaike Information Criterion (AICc) and another on Bayesian Information Criterion (BIC), known to penalize models with a greater number of parameters, particularly BIC, we obtain some competitive models relative do $\Lambda CDM.$ In one of these models the cosmological constant is interpreted as having origin in the creation of matter due to time dependent gravitational field, as opposed to the origin in the vacuum energy.

\end{abstract}
 
 ]




\hspace{1cm} 
\section{Introduction}\label{introduction}
\label{intro}

Since the acceleration of the universe was discovered \cite{riess,perlmutter}, the simplest and most effective explanation to date is to consider the famous cosmological constant created by Einstein almost a century ago \cite{spergel2015dark,einstein2013principle}. Such a constant, originally conceived to describe a static universe \cite{history_Lambda}, is now used to provide the energy necessary to make the universe expand with acceleration in a context of exclusively attractive gravity. Such energy is called dark energy \cite{weight_vacuum,amendola2010dark}.

The simplest explanation for the nature of dark energy is that it is the energy of the vacuum, that is, the intrinsic or fundamental energy of a certain volume of ``empty'' space \cite{weight_vacuum}. This energy corresponds, in the theory of General Relativity, to the effect of the cosmological constant, $\Lambda.$ The observations concerning supernovae leading to the conclusion that the universe is in accelerated expansion are consistent with a very small and positive value for this constant, of the order of $10^{-29}$ $g/cm^3$ \cite{everything_Lambda,rugh2002quantum,straumann1999mystery}.

The problem of the cosmological constant \cite{burgess2013cosmological,straumann1999mystery,weinberg1989cosmological,carroll2001cosmological,padmanabhan2003cosmological} is that the quantum field theories predict a much larger value for this constant, from the calculation of the energy of the quantum vacuum \cite{zel1967cosmological}. In fact, in quantum mechanics, particle and antiparticle pairs are constantly being created from the vacuum, and although these pairs exist for an extremely short time before mutually annihilating each other, this process contributes to the vacuum energy, obtaining a value that, depending on the theory that is used, can be 120 orders of magnitude greater than the value mentioned above and necessary to explain the observations \cite{everything_Lambda}. This is currently seen as one of the fundamental problems of physics and there is currently no solution for it.

Another problem related to the cosmological constant is the so-called problem of cosmic coincidence \cite{cosmic_coincidence_zlatev,cosmic_coincidence_dalal}, which consists in the fact that there is an approximate coincidence between the energy density of the vacuum and the density of matter in the present universe. This is particularly strange given that the relative balance between these energies varies rapidly as the universe expands. In fact, in the primordial universe the energy of the vacuum was negligible in comparison to the matter whereas recently the situation was inverted and it is the energy of the vacuum that began to dominate. There is then a relatively short period in the history of the universe where these energy densities are comparable and it seems a strange coincidence that this period is precisely around the present \cite{coincidence_problem_velten}.

One of the pretensions of this article is to provide a possibility of investigation regarding these problems. The approach chosen is to revisit Rastall's theory of gravitation by providing a formulation of the theory via variational principle \cite{smalley1984variational,smalley1993variational}. Rastall's theory proved resistant to a formulation by a variational principle for about 45 years \cite{rastall1972,smalley1984variational,massa2000einstein}. 

Starting from the proposed Lagrangian density, we obtain the equations of motion and apply them to cosmology. Considering a homogeneous and isotropic universe and using a perfect fluid as constituent material, we obtain the dynamic equations of cosmology. After fitting data for the Hubble function $H(z)$ in terms of the red shift $z$ and using goodness of fit measures based on information theory (corrected Akaike information criterion, AICc) and Bayesian analysis (Bayesian information criterion, BIC), we will see that several of the proposed models fits the data very well, competing with the standard cosmology model with few extra parameters. Some of these models do not consider the cosmological constant as originally conceived. This result encourages us to attack the problem of the cosmological constant from a conventional point of view of the extended theory of relativity, but whose price to be paid is to admit the possibility of creation of matter in large scales originating from the variation of the curvature of space-time.

This article is organized as follows: in section (\ref{Rastall_Theory}) we have a brief review of Rastall's theory followed by a section on its formulation via variational formalism. Section (\ref{cosmology}) establishes the cosmology equations and in section (\ref{fitting}) we use recent data from $H(z)$ to fit several models. The discussion and conclusion in the last two sections conclude the paper.

\section{Rastall's theory of gravitation}
\label{Rastall_Theory}

Consider the Einstein equation of general relativity (GR), $G_{\mu\nu}=\kappa T_{\mu\nu},$ $\kappa$ is a constant, $G_{\mu\nu}\equiv R_{\mu\nu}-1/2g_{\mu\nu}R$ is the Einstein tensor and $T_{\mu\nu}$ the energy-momentum tensor. The Einstein tensor $G_{\mu\nu}$ has the peculiar property that $\nabla^{\nu}G_{\mu\nu}=0,$ as we know from the identities of Bianchi \cite{dirac_GTR}.
This peculiarity of $G_{\mu\nu}$ is unjustifiably extended to the stress-energy tensor, and it is considered to be true that $\nabla^{\nu}T_{\mu\nu}=0.$ But what can be asserted about the covariant divergence of a arbitrary tensor $K_{\mu\nu}$ is that $\nabla^{\nu}K_{\mu\nu}=\lambda a_{\mu},$ $\lambda$ is a constant and $a_{\mu}$ is a vector. 
Several extensions of GR consider the possibility of a violation of $\nabla^{\nu}T_{\mu\nu}=0,$ among which we can mention the $f(R, T)$ \cite{f_R_T} and $f(R,{\cal{L}}_m)$ \cite{harko2014generalized} theories, where $R$ is the Ricci scalar, $T$ is the trace of the energy-momentum tensor, and ${\cal{L}}_m$ is the Lagrangian for matter.

An older theory involving modified gravity violating $\nabla^{\mu}T_{\mu\nu}=0,$ a particular case of the $f(R,T)$ theories mentioned above, is the theory of Rastall \cite{rastall1972}. 
In this theory, a dependency of the covariant divergence of the stress-energy tensor is postulated with the Ricci scalar and is given by $\tensor{T}{^{\mu\nu}_{;\mu}}=\lambda R^{;\nu},$ where $\lambda$ is a constant \cite{rastall1972}. In this way, the field equations in the original formulation of Rastall's theory are
\begin{equation}
R_{\mu\nu}+\left(\kappa\lambda-\frac{1}{2}\right)g_{\mu\nu}R=\kappa T_{\mu\nu}.
\label{feRT}
\end{equation}
General relativity is recovered if $\lambda=0.$

Another useful parameterization of Rastall's model, this time with the cosmological constant $\Lambda$ inserted, is as follows:
\begin{equation}
G_{\mu\nu}=\kappa\left(T_{\mu\nu}-\frac{\gamma-1}{2}g_{\mu\nu}T+\frac{\Lambda}{\kappa}g_{\mu\nu}\right)
\label{Rastall_EoM}
\end{equation}
and
\begin{equation}
\nabla^{\nu}T_{\mu\nu}=\frac{\gamma-1}{2}\nabla_{\mu}T,
\label{Rastall_CoE}
\end{equation}
where $T=g^{\mu\nu}T_{\mu\nu}$ is the trace of the stress-energy tensor and $\gamma=(6\kappa\lambda-1)/(4\kappa\lambda-1).$ 

The cosmological constant was considered as having geometric origin as a result of the identity of Bianchi $$\nabla^{\nu}\left(G_{\mu\nu}-\Lambda g_{\mu\nu}\right)=0.$$ 

One of the criticisms of the model of Rastall is that it does not come from a variational principle. One of the objectives of this study is to provide such a principle. From the proposed Lagrangian emerges an equation for the covariant divergence of the energy-momentum tensor that contains an extra term, not present in the original theory. We investigate the effects of this extra term in the context of cosmology, comparing the results with the conventional Rastall's theory.
We will see that the term involving the cosmological constant should be corrected in a formulation from a variational principle.

\section{The Lagrangian}
\label{variational_formulation}

We propose the following Lagrangian density, composed of the Einstein-Hilbert Lagrangian ${\cal{L}}_{EH}$ and a Rastall Langrangian ${\cal{L}}_R:$
\begin{eqnarray}
\sqrt{-g}{\cal{L}} & = &  \sqrt{-g}({\cal{L}}_{EH}+{\cal{L}}_R)   \nonumber \\
 & \equiv &  \frac{1}{2\kappa}\sqrt{-g}[R+2\Lambda+2\kappa{\cal{L}}_m] \nonumber \\ 
& + & \frac{\alpha}{2}\sqrt{-g}\left[\frac{1}{\kappa}G_{\mu\nu}+ \bar{T}_{\mu\nu}\right]g^{\mu\nu} \nonumber \\
& = & \sqrt{-g}\left[\frac{(1-\alpha)}{2\kappa}R+\frac{\Lambda}{\kappa}+\frac{\alpha}{2} \bar{T}+{\cal{L}}_m\right],
\label{lagrangian}
\end{eqnarray}
where $g$ is the determinant of the metric tensor, ${\cal{L}}_m$ is the Lagrangian density due to matter, $\kappa$ and $\alpha$ are constants and $\bar{T}_{\mu\nu}$ is a arbitrary tensor.
Note that the trace of $\bar{T}_{\mu\nu}$ appears explicitly in the Rastall Lagrangian density and that, curiously, this Lagrangian consists essentially of the equations of motion of general relativity.

In a system of units where $c = G = 1,$ we can write the action as follows:
\begin{equation}
{\cal{A}}=\frac{1}{16\pi}\int f(R,T)\sqrt{-g}d^4x+\int{\cal{L}}_m\sqrt{-g}d^4x
\label{action}
\end{equation}
with $\kappa=8\pi$ and $f(R,T)=(1-\alpha)R+2\Lambda+8\pi\alpha\bar{T}.$ Rastall's theory can be formulated as a particular case of the $f(R,T)$ theory. Formulated in this way, several results for general $f(R,T)$ theory can be used \cite{f_R_T,harko2014thermodynamic}.

Using the well-known facts
\begin{enumerate}[a)]
\item $\delta(\sqrt{-g})=-\frac{1}{2}g_{\mu\nu}\sqrt{-g}\delta g^{\mu\nu},$
\item $\delta(\sqrt{-g}R)=\sqrt{-g}\left(R_{\mu\nu}-\frac{1}{2}Rg_{\mu\nu}\right)\delta g^{\mu\nu},$
\item $\delta(\sqrt{-g}\bar{T})=\sqrt{-g}\left(\bar{T}_{\mu\nu}+\bar{\theta}_{\mu\nu}-\frac{1}{2}\bar{T}g_{\mu\nu}\right)\delta g^{\mu\nu}$\\
with 
$
\bar{\theta}_{\mu\nu}\equiv g^{\alpha\beta}\frac{\delta \bar{T}_{\alpha\beta}}{\delta g^{\mu\nu}}
$
\noindent and
\item $\delta(\sqrt{-g}{\cal{L}}_m)=-\frac{1}{2}T_{\mu\nu}\sqrt{-g}g^{\mu\nu}$
\end{enumerate}
with 
\begin{equation}
T_{\mu\nu}=-\frac{2}{\sqrt{-g}}\frac{\delta(\sqrt{-g}{\cal{L}}_m)}{\delta g^{\mu\nu}},
\label{theta_e_T}
\end{equation}
the field equations are
\begin{equation}
G_{\mu\nu}=\kappa T^{\textrm{(eff)}}_{\mu\nu},
\label{fe1}
\end{equation}
\noindent with  
$$
T^{(\textrm{eff})}_{\mu\nu}\equiv T_{\mu\nu}-\frac{\alpha}{2(1-\alpha)}\left(2\bar{T}_{\mu\nu}+2\bar{\theta}_{\mu\nu}-\bar{T}g_{\mu\nu}\right)+\frac{\Lambda}{\kappa(1-\alpha)}g_{\mu\nu}.
$$
For $\alpha = 0$ we get general relativity.

Considering $\nabla^{\nu}T^{(\textrm{eff})}_{\mu\nu}=0:$
\begin{equation}
\nabla^{\nu}T_{\mu\nu}=\frac{\alpha}{2(\alpha-1)}\left(\nabla_{\mu}\bar{T}-2\nabla^{\nu}\bar{\theta}_{\mu\nu}-2\nabla^{\nu}\bar{T}_{\mu\nu}\right).
\label{nce}
\end{equation}
We see from Eqs.~(\ref{Rastall_CoE}) and (\ref{nce}) that to obtain the theory of Rastall starting from the Lagrangian given by (\ref{lagrangian}), we must have
\begin{enumerate}[i)]
\item $\frac{\alpha}{2(\alpha-1)}=\frac{\gamma-1}{2}$ \hspace{0.3cm} and
\item $\nabla_{\mu}\bar{T}-2\nabla^{\nu}\bar{\theta}_{\mu\nu}-2\nabla^{\nu}\bar{T}_{\mu\nu}=\nabla_{\mu}T,$
\end{enumerate}
which implies
$$
\alpha=\frac{\gamma-1}{\gamma-2}.
$$
and
\begin{equation}
\bar{\theta}_{\mu\nu}=g_{\mu\nu}\bar{\Lambda}-\bar{T}_{\mu\nu}
\label{exigencia}
\end{equation}
respectively, with $2\bar{\Lambda}\equiv\bar{T}-T$ and 
\begin{equation}
\bar{T}_{\mu\nu}=T_{\mu\nu}+2\bar{\Lambda}g_{\mu\nu}.
\label{Tbar}
\end{equation}
Finally, the equations of motion are
\begin{eqnarray}
G_{\mu\nu} & = & \kappa\left[T_{\mu\nu}-\frac{\gamma-1}{2}g_{\mu\nu}T+(2-\gamma)g_{\mu\nu}\frac{\Lambda}{\kappa}\right],
\label{EOM}
\end{eqnarray}
and
\begin{equation}
\nabla^{\nu}T_{\mu\nu}=\frac{\gamma-1}{2}\nabla_{\mu}{T}.
\label{CE}
\end{equation}
Comparing equations (\ref{EOM}) and (\ref{Rastall_EoM}) we see that Rastall's theory in the original version and in the version derived from the variational principle are different with respect to the term with the cosmological constant. In Rastall's theory as proposed here, the cosmological constant as geometry fully on the left side of the field equations can not be interpreted as equivalent to matter-energy (vacuum energy) on the right side of the field equations as is commonly done in general relativity.

To better understand the point in question, we write equation (\ref{EOM}) as follows:
\begin{equation}
G_{\mu\nu}-\Lambda g_{\mu\nu}=\kappa\left[T_{\mu\nu}-\frac{\gamma-1}{2}g_{\mu\nu}T+(1-\gamma)g_{\mu\nu}\frac{\Lambda}{\kappa}\right].
\label{EOM2}
\end{equation}
In Eq.~(\ref{EOM2}) we divide the factor that corresponds to the cosmological constant into two parts: one associated with geometry and another with matter-energy. The degree to which Rastall's theory departs from general relativity is measured by how much $\gamma$ is different from 1, that is, it is measured by the intensity of the cosmological constant due to matter-energy only plus the term corresponding to the trace of the energy-moment tensor. For future discrimination between these two effects, we rewrite Eq.~(\ref{EOM2}) as follows:
\begin{equation}
G_{\mu\nu}-\Lambda g_{\mu\nu}=\kappa\left[T_{\mu\nu}-\frac{\gamma-1}{2}g_{\mu\nu}T+(1-\gamma)g_{\mu\nu}\frac{\hat{\Lambda}}{\kappa}\right],
\label{EOM3}
\end{equation}
where the cosmological constant due to matter-energy creation is indicated by $\hat{\Lambda}.$

\section{Cosmology}
\label{cosmology}

We will consider the homogeneous and isotropic Friedman-Lemaitre-Robertson-Walker (FLRW) universe with scale factor $a,$
\begin{equation}
ds^2=dt^2-a^2(t)\left[\frac{dr^2}{1+Kr^2}+r^2(d\theta^2+\sin^2\theta d\phi^2)\right],
\label{FLRW}
\end{equation}
where $K=1, 0, -1$ is the spatial curvature constant.

For a perfect fluid given by
\begin{equation}
T_{\mu\nu}=(\rho+p)u_{\mu}u_{\nu}-g_{\mu\nu}p
\label{PF}
\end{equation}
with pressure $p,$ density of energy-matter $\rho$ and $u^{\mu}u_{\mu}=1,$ we have $T=\rho-3p$ and the cosmology equations from (\ref{CE}), (\ref{EOM3}) and (\ref{FLRW}) are
\begin{equation}
3\frac{\dot{a}^2}{a^2}=\kappa\left[\rho-\frac{\gamma-1}{2}T\right]+\Lambda-\hat{\Lambda}(\gamma-1)-3\frac{K}{a^2},
\label{F1}
\end{equation}
\begin{equation}
6\frac{\ddot{a}}{a}=2[\Lambda-(\gamma-1)\hat{\Lambda}]-\kappa(6p+\gamma T)
\label{F11}
\end{equation}
Eqs.~(\ref{F1}) and (\ref{F11}) are the Friedmann's and acceleration equations, respectively. 
From Eq.~(\ref{CE}) we have
\begin{equation}
\dot{\rho}+3\frac{\dot{a}}{a}(\rho+p)=\frac{\gamma-1}{2}\dot{T}.
\label{F2}
\end{equation}
If we use the equation of state $p=\omega\rho,$ the cosmological equations (\ref{F1}), (\ref{F11}) and (\ref{F2}) are
\begin{equation}
3\frac{\dot{a}^2}{a^2}=\kappa\rho\left[1-\frac{\gamma-1}{2}(1-3\omega)\right]+\Lambda+\hat{\Lambda}(1-\gamma)-3\frac{K}{a^2},
\label{F111}
\end{equation}
\begin{equation}
6\frac{\ddot{a}}{a}=2[\Lambda-(\gamma-1)\hat{\Lambda}]-\kappa\rho[6\omega+\gamma(1-3\omega)]
\label{F22}
\end{equation}
and
\begin{equation}
\dot{\rho}+3\frac{\dot{a}}{a}(1+\omega)\rho=\frac{\gamma-1}{2}(1-3\omega)\dot{\rho}.
\label{F222}
\end{equation}
The solution of Eq.~(\ref{F222}) is
\begin{equation}
\rho=\rho_0\left(\frac{a_0}{a}\right)^{A}
\label{SF22}
\end{equation}
with $A=\frac{6(1+\omega)}{3-3\omega+\gamma(3\omega-1)},$ $\rho_0$ and $a_0$ are constants. 


Considering dust matter ($\omega=0$), radiation ($\omega=1/3$) and cosmological constant ($\omega=-1$), only for dust matter the expression of $\rho$ is modified by the presence of the constant $\gamma.$ In this case, $\rho\sim a^{-\frac{6}{3-\gamma}}.$


By inserting Eq.~(\ref{SF22}) into Eq.~(\ref{F22}), the acceleration equation becomes
\begin{equation}
\frac{6\ddot{a}}{a}=\kappa\rho_0[3\omega(\gamma-2)-\gamma]a^{-A}+2\Lambda-2(\gamma-1)\hat{\Lambda}.
\label{Friedmann2}
\end{equation}
We see from Eq.~(\ref{Friedmann2}) that it is possible $\ddot{a}>0$ for $\Lambda=\hat{\Lambda}=0.$ The conditions for $\ddot{a}>0$ are:
\begin{equation}
\omega <\frac{1}{3} \hspace{0.3cm} \textrm{and} \hspace{0.3cm} \gamma <\frac{6 \omega }{3 \omega -1}
\label{con1}
\end{equation}
and
\begin{equation}
\omega >\frac{1}{3} \hspace{0.3cm} \textrm{and} \hspace{0.3cm}  \gamma >\frac{6 \omega }{3 \omega -1}.
\label{cond2}
\end{equation}

\subsection{Cosmological parameters}

Observing supernovae, astronomers measure their brightness and redshift. From the brightness we deduce its distance, and from the redshift, the scale factor $a(t)$ at that time. We can express the results of data reduction in terms of the Hubble parameter $H_0$ and the dimensionless parameter $\Omega_0.$ We can write the Friedmann equation according to these conventions. Hubble parameter is defined as 
\begin{equation}
H\equiv\frac{\dot{a}}{a},
\label{HubbleP}
\end{equation}
which is a function of time with present value denoted by $H_0.$ The density parameter is defined as
\begin{equation}
\Omega\equiv\frac{\kappa\rho}{3H^2},
\label{density_parameter}
\end{equation}
which is also a function of time. With $\rho_0$ and $H_0,$ its present value is $\Omega_0.$ $\Omega=1$ corresponds to $K=0,$ $\Omega<1$ corresponds to $K<0$ and $\Omega>1$ corresponds to $K>0.$ The redshift parameter is defined as
\begin{equation}
1+z=y\equiv\frac{a(t_0)}{a(t)}\equiv\frac{1}{a(t)}.
\label{redshift}
\end{equation}
At present, $z=0$ and therefore $y=1.$ It is convenient to set $a$ to unity at the present time, $t_0$ and $a(t_0)\equiv1.$ According to this notation, Friedmann's equation becomes 
\begin{equation}
H^2+\frac{K}{a^2}=H^2\Omega.
\label{cosmo1}
\end{equation}
Because $K$ is a constant, we express it in terms of the present values of the Hubble parameter and the density parameter,
$$
K=H_0^2(\Omega_0-1).
$$
Supposing that $\Omega_0$ pertains to energy in three possible forms, we decompose it into
\begin{equation}
\Omega_0=\Omega_m+\Omega_r+\Omega_{\tilde{\Lambda}},
\label{energy_forms}
\end{equation}
where $\Omega_m$ pertains to matter, $\Omega_r$ pertains to radiation, and $\Omega_{\tilde{\Lambda}}(\equiv\Omega_{\Lambda}+\Omega_{\hat{\Lambda}})$ pertains to vacuum and ``creation of matter'', all present values.

Each component of energy density among these depends on the scale factor $a$ differently: $\rho\propto a^{-A};$ the density $\rho$ as a function of $a$ is expressed as
\begin{eqnarray}
\frac{\kappa}{3}\rho & = & H_0^2\left[\Omega_m a^{-\frac{6}{3-\gamma}}+\Omega_r a^{-4}+\Omega_{\Lambda}+(1-\gamma)\Omega_{\hat{\Lambda}}\right] \nonumber \\
& = & H_0^2\left[\Omega_m y^{\frac{6}{3-\gamma}}+\Omega_r y^4+\Omega_{\Lambda}+(1-\gamma)\Omega_{\hat{\Lambda}}\right]
\label{density0}
\end{eqnarray}
The Friedmann's equation in terms of currently observable parameters, $H_0,$ $\Omega_m,$ $\Omega_r,$ $\Omega_{\Lambda}$ and $\Omega_{\hat{\Lambda}}$ becomes
\begin{eqnarray}
\frac{\dot{a}^2}{a^2} & = & - \frac{H_0^2(\Omega_0-1)}{a^2} \nonumber \\
& + & H_0^2\left[\frac{3-\gamma}{2}\Omega_m a^{-\frac{6}{3-\gamma}}+\Omega_r a^{-4}+\Omega_{\Lambda}+(1-\gamma)\Omega_{\hat{\Lambda}}\right] \nonumber \\
\label{Friedmann0}
\end{eqnarray}
or, equivalently:
\begin{eqnarray}
\left(\frac{H(z)}{H_0}\right)^2 & = & \frac{3-\gamma}{2}\Omega_m(1+z)^{\frac{6}{3-\gamma}} \nonumber \\
& + & \Omega_r(1+z)^4+\Omega_{\Lambda}+(1-\gamma)\Omega_{\hat{\Lambda}} \nonumber \\
& - & (\Omega_0-1)(1+z)^2.
\label{Friedmann1}
\end{eqnarray}

In the next section we will use the Hubble function expressed in Eq.~(\ref{Friedmann1}) to fit the observable parameters using recently obtained experimental data for $H(z).$

\section{Fitting data}
\label{fitting}

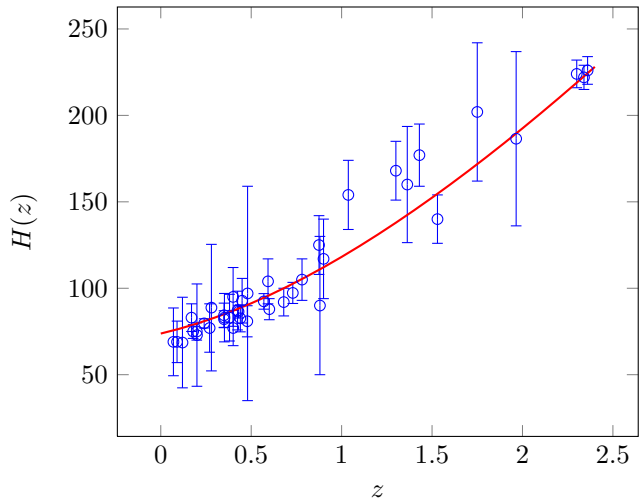
\begin{figure}
\centering
\begin{tikzpicture}[scale = 1.0]
\begin{axis}[domain = 0:2.4, xlabel={$z$}, ylabel={$H(z)$}]]
\addplot [color=blue, only marks, mark=o,]
 plot [error bars/.cd, y dir = both, y explicit]
 table[x =x, y =y, x error=ex, y error =ey,]{H_z_.txt};
\addplot [no markers, red, thick]
{ 
  73.8*sqrt(0.223236*(1+x)^3+(1-0.223236))
};
\end{axis}
\end{tikzpicture}
\caption{$H(z)$ data with error bars and the fitted $\Lambda CDM$ model $H(z)=H_0\sqrt{(1-\Omega)+\Omega(1+z)^{3}},$ with $H_0=73.8$ $km s^{-1} Mpc^{-1}$ \cite{dark_energy_Miao_Li}, $\Omega=0.223,$ $\chi^2_{\textrm{min}}=22.79,$ $AICc=25.11$ and $BIC=26.51.$}
\label{fig1}
\end{figure}

Using data from \cite{DADOS} we fit cosmological parameters present in Eq.~(\ref{Friedmann1}). Such data are shown below and are plotted in Fig.~(\ref{fig1}) along with the fitted curve of the standard model $\Lambda CDM.$


\begin{center}
\tablefirsthead{\toprule $z$ &\multicolumn{1}{c}{$H(z)$} \\ \midrule}
\tablehead{%
\multicolumn{2}{c}%
{{\bfseries  Continued from previous column}} \\
\toprule
$z$ &\multicolumn{1}{c}{$H(z)$}\\ \midrule}
\tabletail{%
\midrule \multicolumn{2}{r}{{Continued on next column}} \\ \midrule}
\tablelasttail{%
\\\midrule
\multicolumn{2}{r}{{Concluded}} \\ \bottomrule}
\begin{supertabular}{cc}
 0.07 & 69  $\pm$ 19.6 \\
 0.09 & 69 $\pm$ 12 \\
 0.12 & 68.6 $\pm$ 26.2 \\
 0.17 & 83  $\pm$ 8 \\
 0.179 & 75  $\pm$ 4 \\
 0.199 & 75  $\pm$ 5 \\
 0.2 & 72.9  $\pm$ 29.6 \\
 0.24 & 79.69  $\pm$ 2.65 \\
 0.27 & 77  $\pm$ 14 \\
 0.28 & 88.8  $\pm$ 36.6 \\
 0.35 & 82.1  $\pm$ 4.9 \\
 0.35 & 84.4  $\pm$ 7 \\
 0.352 & 83  $\pm$ 14 \\
 0.3802 & 83 $\pm$ 13.5 \\
 0.4 & 95  $\pm$ 17 \\
 0.4004 & 77  $\pm$ 10.2 \\
 0.4247 & 87.1  $\pm$ 11.2 \\
 0.43 & 86.45  $\pm$ 3.68 \\
 0.44 & 82.6  $\pm$ 7.8 \\
 0.4497 & 92.8  $\pm$ 12.9 \\
 0.4783 & 80.9  $\pm$ 9 \\
 0.48 & 97  $\pm$ 62 \\
 0.57 & 92.4  $\pm$ 4.5 \\
 0.593 & 104  $\pm$ 13 \\
 0.6 & 87.9  $\pm$ 6.1 \\
 0.68 & 92  $\pm$ 8 \\
 0.73 & 97.3  $\pm$ 7 \\
 0.781 & 105  $\pm$ 12 \\
 0.875 & 125  $\pm$ 17 \\
 0.88 & 90  $\pm$ 40 \\
 0.9 & 117  $\pm$ 23 \\
 1.037 & 154  $\pm$ 20 \\
 1.3 & 168  $\pm$ 17 \\
 1.363 & 160  $\pm$ 33.6 \\
 1.43 & 177  $\pm$ 18 \\
 1.53 & 140  $\pm$ 14 \\
 1.75 & 202  $\pm$ 40 \\
 1.965 & 186.5  $\pm$ 50.4 \\
 2.3 & 224  $\pm$ 8 \\
 2.34 & 222  $\pm$ 7 \\
 2.36 & 226  $\pm$ 8 
\end{supertabular}%
\end{center}







We will fit the $H(z)$ data using some cosmological models proposed below, many of them from equation (\ref{Friedmann1}) and two from purely phenomenological origin proposed in \cite{oscillations1}. These two purely phenomenological models are oscillatory models that best fit the data proposed in \cite{oscillations1,oscillations2}. They are included here for purposes of comparison with the models motivated by equation (\ref{Friedmann1}) and for an update of the conclusions obtained in \cite{oscillations1,oscillations2} with more recent data available. 

Table (\ref{table1}) shows 12 models studied plus $\Lambda CDM$ model considered as benchmark. Of the 12 proposed models, 10 are particular cases of equation (\ref{Friedmann1}) and 2 are the oscillatory models mentioned. For example, the most complex model is the model (H) where $\Omega_m$ represents baryonic matter, $\Omega_K$ represents ``matter associated with the curvature of spacetime,'' $\Omega_r$ is associated with radiation and $\Omega_{\hat{\Lambda}}$ represents the matter associated with the cosmological constant due to matter creation. Another example is the model (L) where we leave $\omega$ as free parameter and we made $\Omega\equiv\Omega_x = 1. $
The table (\ref{table1}) is arranged in order to display the models in ascending order of ``discrepancy from $\chi^2/d.o.f-1$''. This preliminary quality of fit was measured in terms of the $\chi^2$ analysis, where $\chi^2$ is defined by
\begin{equation}
\chi^2(\textrm{parameters})=\displaystyle\sum_{i=1}^{41}\frac{[H_{\textrm{mod}}(\textrm{parameters};z_i)-H_{\textrm{obs}}(z_i)]^2}{\sigma^2(z_i)},
\label{chi}
\end{equation}
where $H_{\textrm{mod}}$ is the predicted value for the Hubble parameter in the assumed model, $H_{\textrm{obs}}$ is the observed value, $\sigma$ corresponds to $1\sigma$ uncertainty, and the summation is over the 41 observational $H(z)$ data points at redshift $z_i.$ The parameters are estimated by the minimization of $\chi^2.$ What actually orders the models with respect to their quality of fit is $\chi^2$ divided by the degrees of freedom (d.o.f). Values  of $\chi^2/d.o.f$ close to 1 ($\chi^2/d.o.f\approx 1$) represent better fittings \cite{error_taylor}.

All models in the table (\ref{table1}) that do not have $\Lambda CDM$ in their names are models based on Rastall's theory and therefore in equation (\ref{Friedmann1}). This fact is evidenced by the presence of $\gamma$ in the equations that define them. In the next section we will consider better strategies for selecting models based on information criteria and Bayesian analysis.

\subsection{Model selection}

Some model selection methods are defined in terms of an appropriate \emph{information criterion}, a mechanism well-founded theoretically that uses data to give a model a score, which leads to a ranked list of candidate models from the best to the worst \cite{burnham_anderson_2004,burnham_anderson_book,liddle2007information}. Two of the most important of these information criteria are the Akaike information criterion (AIC) and Bayesian information criterion (BIC). The general formulas for models with vector parameter $\hat{\theta}$ are
\begin{equation}
AIC=-2\log{{\cal{L}}(\hat{\theta}|\textrm{data})}+2p
\label{AIC}
\end{equation}
and
\begin{equation}
BIC=-2\log{{\cal{L}}(\hat{\theta}|\textrm{data})}+\log{(n)}p,
\label{BIC}
\end{equation}
where $\log{{\cal{L}}}(\hat{\theta}|\textrm{data})$ is the log-likelihood of the model, $p$ is the number of parameters and $n$ is the sample size. For models with normally distributed residuals, $\log{{\cal{L}}}=-\frac{n}{2}\log{\chi^2}$ with $\chi^2$ given by Eq.~(\ref{chi}).

The AIC and BIC criteria act as a penalized log-likelihood criterion, providing a balance between good fit (high value of log-likelihood) and complexity (complex models are penalized more that simple ones). These criteria punishes the models for being too complex in the sense of containing many parameters. The models with the lowest AIC and BIC scores are selected.

It is useful to briefly mention how these criteria behave in relation to their properties of \emph{consistency} and \emph{efficiency}. The comparison is based in the study of the \emph{penalty} applied to the maximized log-likelihood value in a framework with \emph{increasing sample size}. If we make the assumption that there are one \emph{true model} that generates the data and that this model \emph{is one of the candidate models}, we would want the model selection method to identify this true model. This is related to consistency. From \cite{model_selection_Hjort} (our emphasis):
\begin{quotation}
A model selection method is weakly consistent if, with probability tending to one \emph{as the sample size tends to infinity}, the selection method is able to select the true model from the candidate models. Strong consistency is obtained when the selection of the true model happens almost surely.
\end{quotation}
When we do not assume that the true model is present among the proposed models, as we believe to be the case in most practical situations, including cosmology, we can assume that a candidate model is the closest to the true model in the Kullback-Leibler distance sense (see bellow) \cite{burnham_anderson_book}. In this case \cite{model_selection_Hjort},
\begin{quotation}
\noindent $\cdots$ we can state weak consistency as the property that, with probability tending to one [\emph{as the sample size tends to infinity}], the model selection method picks such a closest model.
\end{quotation}
From the foregoing we see that a strongly desirable condition for the use of a consistent criterion is to have a sufficiently large sample size. This fact, among others that will be mentioned below, makes us less likely to take the results very strictly for BIC, at least with the amount of data currently available for $H(z).$

Another desirable property of selection methods, efficiency, can be roughly described by \cite{model_selection_Hjort}:
\begin{quotation}
$\cdots$ [W]e might want an information criterion to posses is that it behaves ``almost as well'', in terms of mean square error, or squared error loss.
\end{quotation}
In \cite{model_selection_Hjort} it is proved that
\begin{quotation}
$\cdots$ AIC is not strongly consistent, though it is efficient, while the opposite is true for the BIC.
\end{quotation}
Given the above assertion, it is useful to question whether a model selection method that contemplates both desirable properties, BIC consistency with AIC efficiency, exists. The answer is \emph{not}, as proven in \cite{yang2005can}. This means that when selecting a method to choose the best models, we are necessarily making a choice between consistency and efficiency. The optimal or most convenient choice is conditioned to the size of the sample $n$ as well as to the \emph{world view} of the researcher, as will be discussed in the section (\ref{discussion}).

We have seen that for BIC, the properties of consistency are defined (rigorously in \cite{model_selection_Hjort}) in terms of asymptotic properties related to sample size $n.$ If this size is not ``large enough'', the practical validity of the theorems weaken. 

For AIC something analogous occurs. In its derivation, an approximation is made that is valid for large sample sizes. A rule of thumb was proposed in \cite{burnham_anderson_book} to define what would be a ``sample size not large enough''. Such a rule states that if $n /p\leq 40,$ the results for the AIC method may be less accurate. To mitigate this problem, a correction was proposed for the Akaike method, the so-called \emph{corrected Akaike information criterion}, AICc, whose formula is given, under certain assumptions, by \cite{burnham_anderson_book}
\begin{equation}
AICc=AIC+\frac{2(p+1)(p+2)}{n-p-2}.
\label{AICc}
\end{equation}
AICc is essentially AIC with a greater penalty for extra parameters. Using AIC, instead of AICc, when $n$ is not many times larger than $p^2,$ increases the probability of selecting models that have too many parameters, \textit{i.e.}, of overfitting.
Due to the fact that we are using data with $n = 41,$ we will use AICc as the criterion for selecting models based on Akaike.

AIC, AICc and BIC (collectively dubbed IC, from Information Criteria) are absolute numbers that have no significance when evaluated in isolation. What matters is the value of the difference of IC from two different models. In order to better characterize this fact, it is common to calculate $\Delta_i IC \equiv IC_i-IC_ *,$ where we calculate the difference of IC values for two models, $i$ and $*,$ where the model $*$ is established as the reference model. In this work we consider as reference model the standard model of cosmology, the $\Lambda CDM$ model. These $\Delta_i IC$ are easy to interpret and allow a quick strength-of-evidence comparison of candidate models. According to \cite{burnham_anderson_2004}:
\begin{quotation}
Some simple rules of thumb are often useful in assessing the relative merits of models in the set: Models having $\Delta_i\leq 2$ have substantial support (evidence), those in which $4\leq\Delta_i\leq7$ have considerably less support, and models having $\Delta_i>10$ have essentially no support.
\end{quotation}

Another useful tool for providing weights of evidence for each of the $R=11$ (phenomenological models were excluded) models considered in the analysis are the Akaike weights $w_i$ given by \cite{burnham_anderson_2004}
\begin{equation}
w_i=\frac{e^{-\frac{\Delta_i}{2}}}{\displaystyle\sum_{r=1}^Re^{-\frac{\Delta_r}{2}}}.
\label{pesos}
\end{equation}
where $\Delta_i$ represents $\Delta_i AICc$ or $\Delta_i BIC.$
The $w_i$ from AICc can be interpreted as the probability of model $i$ being, in fact, the best model in the sense of the Kulback-Leibler's (K-L) distance \cite{burnham_anderson_2004}. K-L information $I(f,g)$ is the \emph{information lost} when model $g(x,\theta)$ is used to approximate the ``full reality or truth'', $f;$ this is defined for continuous functions as the integral
\begin{equation}
I(f,g)=\int f(x)\log{\left(\frac{f(x)}{g(x|\theta)}\right)dx}.
\label{K-L}
\end{equation}
The best model loses the minimum amount of information possible and Akaike's criterion seeks precisely this by minimizing the distance of K-L.

All our data analysis results are shown in tables (\ref{table1}), (\ref{table2}) and (\ref{table3}), the latter in the appendix, and in figure (\ref{fig2}). Table (\ref{table1}) shows all models with their parametrizations, table (\ref{table2}) presents the AICc and BIC measurements and their variants, and figure (\ref{fig2}) presents a column chart to facilitate the general understanding of the hierarchy of models with regard to the rule of thumb. For the calculation of $w_i$ in table (\ref{table2}) we do not consider the weights relative to the phenomenological models (B) and (C). In the next section we will discuss the results.

\begin{table*}[!htb]
  \centering
  \begin{tabular}{*{3}{c}}
\hline
\hline
 \textbf{Model} & \pmb{$\left(H(z)/H_0\right)^2$}, $H_0=73,8$ $km~s^{-1}~Mpc^{-1}$ \cite{dark_energy_Miao_Li} & \pmb{$\chi^2/d.o.f$}\\
\hline
\hline
$\Omega_m+\Omega_r$ & $\frac{1}{2} (3-\gamma ) \Omega_m  (z+1)^{\frac{6}{3-\gamma }}+(1-\Omega_m ) (z+1)^4$  & 1.764 (A)\\
$\Lambda CDM$ Oscillation& $ a_1 \cos \left(a_2 z^2+a_3\right)+1-a_1 \cos (a_3)-\Omega_m+\Omega_m  (z+1)^3$ & 0.3388 (B)\\
Oscillation & $a_1\cos\left(a_2 z^2+a_3\right)+1-a_1\cos(a_3)-\Omega_m+\frac{1}{2}(3-\gamma)\Omega_m  (z+1)^{\frac{6}{3-\gamma}}$ & 0.3482 (C) \\
$\Omega_m+\Omega_{\Lambda}+\Omega_r$ & $1-\Omega_m -\Omega_r+\frac{1}{2} (3-\gamma ) \Omega_m  (z+1)^{\frac{6}{3-\gamma }}+\Omega_r (z+1)^4$ & 0.4726 (D)\\
$\Omega_m+\Omega_{\Lambda}+\Omega_{\hat{\Lambda}}$ & $1-\Omega_{\Lambda} -\Omega_m+(1-\gamma) \Omega_{\Lambda} +\frac{1}{2} (3-\gamma ) \Omega_m  (z+1)^{\frac{6}{3-\gamma }}$ & 0.4876 (E)\\
$\Omega_m+\Omega_{\hat{\Lambda}}+\Omega_r$ & $(1-\gamma) (1-\Omega_m -\Omega_r)+\frac{1}{2} (3-\gamma ) \Omega_m  (z+1)^{\frac{6}{3-\gamma }}+\Omega_r (z+1)^4$ & 0.5032 (F)\\
$\Omega_m+\Omega_{\Lambda}$ & $1-\Omega_m +\frac{1}{2} (3-\gamma ) \Omega_m  (z+1)^{\frac{6}{3-\gamma }}$ & 0.5120 (G)\\
$\Omega_m+\Omega_K+\Omega_r+\Omega_{\hat{\Lambda}}$ & $(1-\gamma) (1-\Omega -\Omega_r)+(K-1) (z+1)^2+\frac{1}{2} (3-\gamma ) \Omega_m  (z+1)^{\frac{6}{3-\gamma }}+\Omega_r (z+1)^4$ & 0.5168 (H)\\
$\Lambda CDM+\Omega_r$ & $1-\Omega_m -\Omega_r+\Omega_m  (z+1)^3+\Omega_r (z+1)^4$ & 0.5312 (I)\\
$\Omega_m+\Omega_K+\Omega_r$ & $(K-1) (z+1)^2+\frac{1}{2} (3-\gamma ) \Omega_m  (z+1)^{\frac{6}{3-\gamma }}+(1-\Omega_m ) (z+1)^4$ & 0.5411 (J)\\
\rowcolor{Gray}
\pmb{$\Lambda CDM$} &  \pmb{$(1-\Omega)+\Omega(z+1)^3$}  & \pmb{0.5698} \pmb{($*$)}\\
$\Omega_m+\Omega_{\hat{\Lambda}}$ & $(1-\gamma) (1-\Omega_m )+\frac{1}{2} (3-\gamma ) \Omega_m  (z+1)^{\frac{6}{3-\gamma }}$ & 0.5769 (K)\\
$\Omega_x=1,$ $\omega$ & $\left[1-\frac{1}{2} (\gamma -1) (1-3 \omega )\right] (z+1)^{\frac{6 (\omega +1)}{\gamma  (3 \omega -1)-3 \omega +3}}$ & 0.7379 (L)\\
\hline 
\hline
\end{tabular}\caption{All functions $H(z)$ fitted, ordered in terms of $\chi^2/d.o.f$ from (A) to (L), from $\chi^2/d.o.f$ furthest from 1 to $\chi^2/d.o.f$ closer to 1, with $\Lambda CDM,$ our benchmark, emphasized. All models without $\Lambda CDM$ in front of the name refer to the models that originated in Rastall's theory, as verified by the presence of $\gamma$ in their respective equations.}
\label{table1}
\end{table*}

\begin{table*}[!htb]
  \centering
  \begin{tabular}{*{8}{c}}
\hline
\hline
 \textbf{Model} & \pmb{$\chi^2_{\textrm{min}}$} & AICc & BIC & $\Delta AICc$ & $\Delta BIC$ & $w_i AICc$ & $ w_i BIC$ \\
\hline
\hline
$\Omega_m+\Omega_r$   & 68.80 (A) & 73.12 (13) & 76.22 (13) & 48.01 & 49.71 & 0.00 & 0.00\\
$\Omega_x=1,$ $\omega$ & 28.78 (L) & 33.10 (12) & 36.21 (12) & 7.99 & 9.70 & 0.00 & 0.00\\
\rowcolor{Gray}
\pmb{$\Lambda CDM$}  & \pmb{22.79} \pmb{($*$)} & \pmb{25.11 (7)} & \pmb{26.51 (1)} & \pmb{0} & \pmb{0} & 0.13 & 0.34\\
\rowcolor{Cyan}
$\Omega_m+\Omega_{\hat{\Lambda}}$ & 22.50 (K) & 26.82 (9) & 29.93 (7) & 1.71 & 3.42 & 0.05 & 0.06\\
\rowcolor{Red1}
$\Lambda CDM+\Omega_r$ & 20.72 (I) & 25.04 (6) & 28.14 (4) & -0.07 & 1.63 & 0.13 & 0.15\\
\rowcolor{Cyan}
$\Omega_m+\Omega_K+\Omega_r$ & 20.56 (J) & 26.88 (10) & 31.70 (10) & 1.77 & 5.19 & 0.05 & 0.03\\
\rowcolor{Red1}
\pmb{$\Omega_m+\Omega_{\Lambda}$} & \pmb{19.97 (G)} & \pmb{24.29 (3)} & \pmb{27.40 (3)} & \pmb{-0.82} & \pmb{0.89} & 0.19 & 0.22\\
\rowcolor{Cyan}
$\Omega_m+\Omega_K+\Omega_r+\Omega_{\hat{\Lambda}}$ & 19.13 (H) & 27.44 (11) & 33.98 (11) & 2.33 & 7.47 & 0.04 & 0.01\\
\rowcolor{Cyan}
$\Omega_m+\Omega_{\hat{\Lambda}}+\Omega_r$ & 19.13 (F) & 25.44 (8) & 30.26 (8) & 0.33 & 3.75 & 0.11 & 0.05\\
\rowcolor{Cyan}
$\Omega_m+\Omega_{\Lambda}+\Omega_{\hat{\Lambda}}$ & 18.53 (E) & 24.85 (5) & 29.67 (6) & -0.26 & 3.16 & 0.14 & 0.07\\
\rowcolor{Cyan}
$\Omega_m+\Omega_{\Lambda}+\Omega_r$ & 18.43 (D) & 24.75 (4) & 29.57 (5) & -0.36 & 3.06 & 0.15 & 0.07\\
\rowcolor{Red1}
\pmb{$\Lambda CDM$ Oscillation} & \pmb{12.54} (B) & \pmb{20,86} (1) & \pmb{27.39} (2) & \pmb{-4.25} & \pmb{0.88} & $-$ & $-$\\
\rowcolor{Cyan}
Oscillation & 12.54 (C) & 22.86 (2) & 31.1 (9) & -2.25 & 4.59 & $-$ & $-$\\
\hline 
\hline
\end{tabular}
\caption{All functions $H(z)$ fitted, ordered in terms of $\chi^2_{\textrm{min}},$ showing the valus of $AICc,$ $BIC,$ $\Delta AICc,$ $\Delta BIC,$ $w_i AICc$ and $w_i BIC.$ Models that pass in the BIC criterion are evidenced in orange color and models that pass in the AICc criterion are evidenced in light cyan. Phenomenological models were excluded in the calculations of $w_i.$}
\label{table2}
\end{table*}

\section{Discussion}
\label{discussion}

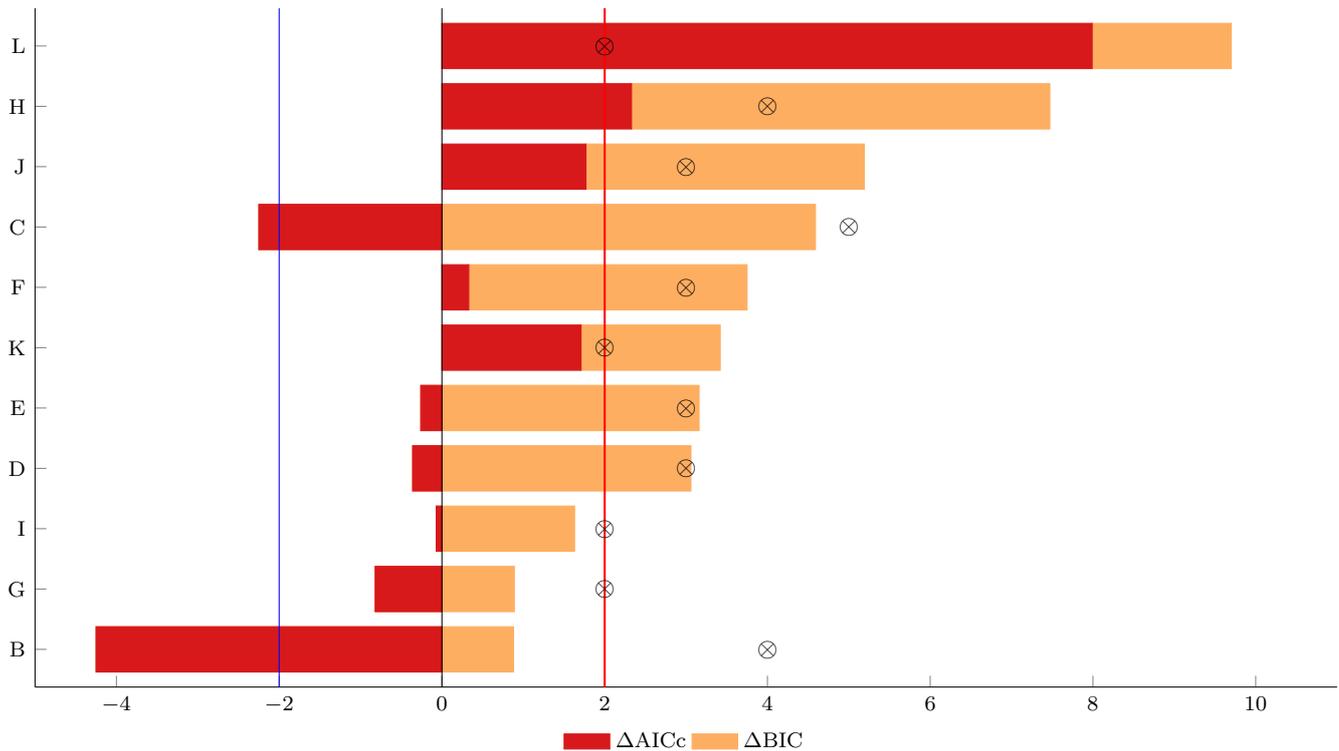
\begin{figure}[!htb]
\centering
\begin{tikzpicture}
\begin{axis}[
    xbar stacked,
    legend style={
    legend columns=2,
        at={(xticklabel cs:0.5)},
        anchor=north,
        draw=none
    },
    ytick=data,
    axis y line*=none,
    axis x line*=bottom,
    tick label style={font=\footnotesize},
    legend style={font=\footnotesize},
    label style={font=\footnotesize},
    xtick={-4,-2,0,2,4,6,8,10},
    width=1.0\textwidth,
    bar width=6mm,
    xlabel={Time in ms},
    yticklabels={B, G, I, D, E, K, F, C, J, H, L},
    xmin=-5,
    xmax=11,
    area legend,
    y=8mm,
    enlarge y limits={abs=0.625},
]
\draw[red, thick] (axis cs: 2,\pgfkeysvalueof{/pgfplots/ymin}) -- 
                      (axis cs: 2,\pgfkeysvalueof{/pgfplots/ymax})node[anchor=west,rotate=90]{Some label};
\draw[blue] (axis cs: -2,\pgfkeysvalueof{/pgfplots/ymin}) -- 
                      (axis cs: -2,\pgfkeysvalueof{/pgfplots/ymax})node[anchor=west,rotate=90]{Some label};
\draw[black] (axis cs: 0,\pgfkeysvalueof{/pgfplots/ymin}) -- 
                      (axis cs: 0,\pgfkeysvalueof{/pgfplots/ymax})node[anchor=west,rotate=90]{Some label};
\addplot[findOptimalPartition,fill=findOptimalPartition] coordinates
{(-4.25,0) (-0.82,1) (-0.07,2) (-0.36,3) (-0.26,4) (1.71,5) (0.33,6) (-2.25,7) (1.77,8) (2.33,9) (7.99,10)};
\addplot[storeClusterComponent,fill=storeClusterComponent] coordinates
{(5.13,0) (1.71,1) (1.7,2) (3.42,3) (3.42,4) (1.71,5) (3.75-0.33,6) (4.59+2.25,7) (5.19-1.77,8) (7.47-2.33,9) (9.7-7.99,10)};
\legend{$\Delta$AICc, $\Delta$BIC}
\coordinate (A) at (90,0);
\coordinate (B) at (70,10);
\coordinate (C) at (70,20);
\coordinate (D) at (80,30);
\coordinate (E) at (80,40);
\coordinate (F) at (70,50);
\coordinate (G) at (80,60);
\coordinate (H) at (100,70);
\coordinate (I) at (80,80);
\coordinate (J) at (90,90);
\coordinate (K) at (70,100);
\end{axis}  
\node at (A) {$\otimes$};
\node at (B) {$\otimes$};
\node at (C) {$\otimes$};
\node at (D) {$\otimes$};
\node at (E) {$\otimes$};
\node at (F) {$\otimes$};
\node at (G) {$\otimes$};
\node at (H) {$\otimes$};
\node at (I) {$\otimes$};
\node at (J) {$\otimes$};
\node at (K) {$\otimes$};
\end{tikzpicture}
\caption{Model selection bar chart. The bars are ordered according to the classification by $\Delta$BIC in table (\ref{table2}). Model (A) is not present due to poor fit distorting the horizontal scale. The red vertical line represents the threshold above which the models do not show statistical support according to the standard rule of thumb. Analogous reasoning holds for the blue line, with the rule \emph{against} the standard model. Number of parameters of each model (collected from horizontal axis) are represented by $\otimes.$}
\label{fig2}
\end{figure}

The most important result can be seen in Fig.~(\ref{fig2}) where almost all proposed models pass the test when we consider the AICc criterion.

\begin{figure}
  \centering
    \includegraphics[width=1.0\textwidth]{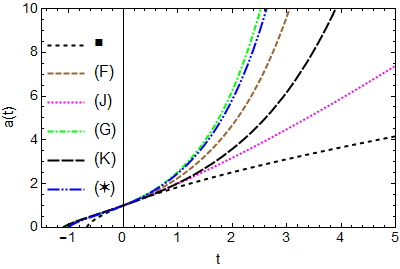}
      \caption{Plot of $a(t)$ for selected models. The square $\blacksquare$ represents a model only with matter without cosmological constant: $a'(t)=H_0\sqrt{\frac{1}{a(t)}}.$ All models use $H_0=1.$}
  \label{a_t}
\end{figure}

The most demanding criterion, BIC, disapproves almost all, except the models (B), (G) and (I). Model (B) represents one of the phenomenological models, which by the way did very well again with the most recent $ H(z) $ data. The model (I) is very similar to the standard model, differing only by the inclusion of radiation. Some curves for $a(t)$ for some models are shown in Fig. (\ref{a_t}), where the differential equations were solved numerically using the data of table (\ref{table3}).

The model (G), a very good model according to both AICc and BIC, is analogous to the standard $\Lambda CDM$ model but incorporates the constant $\gamma \neq 1$ (see table \ref{table3}). Consider it seriously as well as all models with $ \gamma \neq 1 $ implies the real possibility of matter being created with the temporal variation of the gravitational field \cite{harko2014thermodynamic}. Exactly because it is analogous to the standard model, it has the drawback of inheriting all its conceptual problems related to the cosmological constant and also adding the problem of the observational verification of the creation of matter.

Models approved by the rule of thumb with BIC tend to be very restrictive. We will now consider some interesting possibilities involving AICc. Further discussion as to the appropriateness of these criteria will be made at the conclusion.

Models considered competitive relative to $ \Lambda CDM $ but that differ significantly from it (model (I) for example is only $ \Lambda CDM $ with radiation) are models (D), (E), (K), (F) and (J). All these models satisfies $\Delta AICc\leq 2.$ Model (D) is the Rastall's version for the model $\Lambda CDM $ with radiation [model (I)], and it better describes the observed data according to AICc. An undesirable property (in principle) of this model when adjusted to the data used is its negative value for $ \Omega_r $ and its large standard error (see table (\ref{table3})). The use of more complete data other than $ H(z) $ may shed some light on this.

In general, all models suffer from a high standard error in determining the constant $ \gamma, $ except model (K) where the problem is less severe.
Fig.~(\ref{a_t}) shows that such a model predicts a slower expansion for $a(t)$ when compared to $ \Lambda CDM.$ This model also presents an excellent value for the root mean square error (RMSE) \cite{RMSE}. In addition, this model is conceptually particularly attractive because it involves only the cosmological constant that originates in the creation of matter, $\hat{\Lambda}.$ If we could disassociate the accelerated expansion of the universe with a cosmological constant that originates from the energy of the vacuum, naturally some problems related to such a constant would be alleviated, if not fully resolved. Of course in this case we would only be switching the problem since we would have to observe the creation of matter. But perhaps this can be a fruitful exchange, since the problem of the cosmological constant has been resistant to the solution for several decades. 

An interesting parallel can be drawn between the model (G), which is the manifestation \textit{a la} Rastall of the standard model, and the model (K), the best manifestation of our desire to get away from the problem of the cosmological constant. It is not possible to obtain the model (K) starting from Rastall's theory as originally formulated. Model (G) is the closest model (from the standpoint of the standard model $\Lambda CDM$) we would get that originates from Rastall's original theory, not obtained from a variational principle. The model (K) is the model that best characterizes the original aspect of Rastall's theory as proposed here, originated from a variational principle. We conclude that what allowed us to explain the accelerated expansion of the universe without using a cosmological constant from vacuum energy was the variational formulation of Rastall's theory.

\section{Conclusion}
\label{conclusion}

In this paper we propose a formulation of Rastall's theory based on a variational principle. This task, simplified by recent theoretical developments in alternative models of general relativity, has allowed the proposal of a hierarchy of models not yet sufficiently studied. This hierarchy of models was built using model selection techniques based on information theory and Bayesian analysis. These criteria are complementary, although they have conflicting properties and it is difficult to establish which one should be preferred. Preference criteria in this choice are subjective and closely connected with the modeler's world view.
Akaike information criterion is a class of model \emph{selection} tools that provide the best predictive accuracy. Bayesian information criterion is a class of \emph{confirmation/falsification} tools that are consistent. Which of them should be used? 

When we mention two of the desirable properties of any model classification method, namely, efficiency and consistency, we have seen that such properties are excludents: a method that incorporates both is not possible. This state of affairs is best understood intuitively if we conceive the fact that such properties correspond to different world views. Let's characterize these world views as follows:
\begin{enumerate}[i)]
\item Researchers with world view \textcircled{a} believe that a very complex model, perhaps inaccessible to complete human understanding, produces the data. Because of this unknowability, they assume that $ p\gg n$ and do not expect candidate models to correspond exactly to reality. The most they can hope for is \emph{selecting the model} for better forecasting;
\item For researchers with world view \textcircled{b}, a relatively simple process, whose underlying objective reality is accessible, produces the data. The sample size of data, $n,$ greatly exceeds the model parameter space ($p\ll n$). One of the candidate models fitted to the data is actually equivalent to the true model that produced the data and so \emph{is} the objective reality. The task of these researchers is more associated with model confirmation/falsification.
\end{enumerate}
The model that best fits the data must be interpreted in different ways according to the \textcircled{a} or \textcircled{b} view assumed, implicitly or explicitly. In vision \textcircled{a}, we will never find the truth, we can only find the model that maximizes predictive accuracy. In vision \textcircled{b}, we really expect to find the correct model that describes objective reality as sample size grows. Akaike methods are appropriate for situations analogous to the world view \textcircled{a}, and Bayesian methods are appropriate for situations similar to the world view \textcircled{b}. It is the view of the authors of this paper, that the history and philosophy of science seem to support, that view \textcircled{a} is more appropriate and realistic in physics and cosmology. Because of this we do not believe it is prudent to restrict the possibilities too much using BIC. This view stimulates and encourages us to further study the properties of models that contemplate the possibility of explaining the accelerated expansion of the universe via temporal variation of the gravitational field and the corresponding creation of matter, as in model (K).

\appendix

\section{Appendix}

In this appendix we present in table (\ref{table3}) all the models fitted with the values of their respective parameters as well as the value of the root mean square RMSE.

\begin{table*}[!htb]
  \centering
  \begin{tabular}{*{6}{c}}
\hline
\hline
\textbf{Model} & \textbf{Parameters} & Values & Standard error & Confidence Interval & RMSE \\
\hline
\hline
(A) & $\Omega_m, \gamma$   & $\Omega_m=2.52326$ & $0.0456223$ & $\{2.47704, 2.56947\}$ & $13.61$ \\
&                      & $\gamma=1.42942$ &   $0.00469647$ & $\{1.42466, 1.43417\}$ &  \\
(L) & $\gamma,$ $\omega$   & $\gamma=1.27113$ &  $0.0189034$ & $\{1.25198, 1.29028\}$ & $10.79$ \\
&                      & $\omega=-0.527717$ & $0.0123655$ & $\{-0.540243, -0.515191\}$ & \\                      
\rowcolor{Gray}
\pmb{$(*)$} &\pmb{$\Omega$}  & \pmb{$\Omega=0.223236$} & \pmb{$0.00873264$} & \pmb{$\{0.214393, 0.23208\}$} & \pmb{12.10} \\
(K) &$\Omega_m, \gamma$   & $\Omega_m=0.377151$ & $0.0113577$ & $\{0.365646, 0.388657\}$ & $10.84$ \\
&                      & $\gamma=0.541202$ &   $0.0389177$ & $\{0.501778, 0.580626\}$ &  \\
(I) &$\Omega_m,$ $\Omega_r$   & $\Omega_m=0.160909$ &  $0.0437373$ & $\{0.116603, 0.205216\}$ & $12.71$ \\
&                      & $\Omega_r=0.0202814$ & $0.0139837$ & $\{0.00611585, 0.034447\}$ & \\ 
(J) &$\Omega_m, \gamma, K$   & $\Omega_m=0.941743$ &  $0.0157855$ & $\{0.925747, 0.957739\}$ & $12.31$ \\
&                      & $K=0.0521157$ & $0.0614141$ & $\{-0.0101186, 0.11435\}$ & \\ 
&                 & $\gamma=-0.60197$ & $0.278139$ & $\{-0.883823, -0.320116\}$ & \\ 
(G) &$\Omega_m, \gamma$   & $\Omega_m=0.187023$ & $0.0237533$ & $\{0.162961, 0.211086\}$ & $12.54$ \\
&                      & $\gamma=1.1427$ &   $0.0867217$ & $\{1.05485, 1.23055\}$ &  \\
(H) &$\Omega_m, \gamma, K, \Omega_r$   & $\Omega_m=0.249322$ & $284.314$ & $\{-287.966, 288.464\}$ & $11.88$ \\
&                      & $\gamma=0.290255$ &   $291.403$ & $\{-295.111, 295.692\}$ &  \\
&                      & $K=0.988712$ &   $428.966$ & $\{-433.862, 435.84\}$ &  \\
&                      & $\Omega_r=0.0324707$ &   $0.729464$ & $\{-0.707001, 0.771943\}$ &  \\
(F) &$\Omega_m, \gamma, \Omega_r$   & $\Omega_m=0.24182$ & $0.0731334$ & $\{0.16771, 0.31593\}$ & $11.89$ \\
&                      & $\gamma=0.297836$ &   $0.135457$ & $\{0.16057, 0.435101\}$ &  \\
&                      & $\Omega_r=0.0324613$ &   $0.0144455$ & $\{0.0178229, 0.0470998\}$ &  \\
(E) &$\Omega_m, \gamma, \Omega_{\hat{\Lambda}}$   & $\Omega_m=0.245662$ & $0.0609698$ & $\{0.183878, 0.307446\}$ & $11.68$ \\
&                      & $\gamma=0.95634$ &   $0.194274$ & $\{0.759471, 1.15321\}$ &  \\
&    & $\Omega_{\hat{\Lambda}}=0.11527$ &   $0.128262$ & $\{-0.0147047, 0.245245\}$ &  \\
(D) &$\Omega_m, \gamma, \Omega_{r}$   & $\Omega_m=0.52637$ & $0.282041$ & $\{0.240562, 0.812178\}$ & $11.28$ \\
&                      & $\gamma=1.37401$ &   $0.0547329$ & $\{1.31854, 1.42947\}$ &  \\
&    & $\Omega_{r}=-0.225615$ &   $0.188242$ & $\{-0.41637, -0.0348587\}$ &  \\
(B) & $\Omega_m, a_1, a_2, a_3$   & $\Omega_m=0.230154$ & $0.0112571$ & $\{0.218743, 0.241566\}$ & $10.00$ \\
&                      & $a_1=-0.689671$ &   $0.265715$ & $\{-0.959032, -0.420311\}$ &  \\
&                      & $a_2=-2.48397$ &   $0.239258$ & $\{-2.72652, -2.24143\}$ &  \\
&                      & $a_3=7.01061$ &   $0.183163$ & $\{6.82493, 7.19629\}$ &  \\
(C) & $\Omega, a_1, a_2, a_3, \gamma$  & $\Omega_m=0.231265$ & $0.065034$ & $\{0.165314, 0.297216\}$ & $10.00$ \\
&                      & $a_1=0.690777$ &   $0.277451$ & $\{0.409412, 0.972141\}$ &  \\
&                      & $a_2=2.48847$ &   $0.319206$ & $\{2.16476, 2.81218\}$ &  \\
&                      & $a_3=2.40863$ &   $0.353219$ & $\{2.05042, 2.76683\}$ &  \\
&                   & $\gamma=0.996315$ &   $0.211654$ & $\{0.781675, 1.21095\}$ &  \\
\hline 
\hline
\end{tabular}
\caption{Parameters of the $H(z)$ fitted. RMSE is the Root Mean Square Error \cite{RMSE}.}
\label{table3}
\end{table*}


\bibliographystyle{pnas.bst}
\bibliography{Relatividade.bib}

\end{document}